\documentclass{article}
\usepackage{spconf,amsmath,graphicx,booktabs,xspace,amsfonts,cite,multirow,makecell,amsthm,amssymb,mathrsfs,bm}

\title{Multiscale Matching Driven by Cross-Modal Similarity Consistency \\ for Audio-Text Retrieval}

\name{Qian Wang, Jia-Chen Gu, Zhen-Hua Ling$^{\ast}$ \thanks{*Corresponding author}}

\address{National Engineering Research Center of Speech and Language Information Processing,\\University of Science and Technology of China, Hefei, P.R.China\\
 {\tt wangq621@mail.ustc.edu.cn, \{gujc,zhling\}@ustc.edu.cn}}

\begin{document}
\ninept
\maketitle
\begin{abstract}
Audio-text retrieval (ATR), which retrieves a relevant caption given an audio clip (A2T) and vice versa (T2A), has recently attracted much research attention. 
Existing methods typically aggregate information from each modality into a single vector for matching, but this sacrifices local details and can hardly capture intricate relationships within and between modalities. Furthermore, current  ATR datasets lack comprehensive alignment information, and simple binary contrastive learning labels overlook the measurement of fine-grained semantic differences between samples. To counter these challenges, we present a novel ATR framework that comprehensively  captures the matching relationships of multimodal information from different perspectives and finer granularities. Specifically, a fine-grained alignment method is introduced, achieving a more detail-oriented matching through a multiscale  process from local to global  levels to capture meticulous cross-modal relationships. In addition,  we pioneer the application of cross-modal similarity consistency, leveraging intra-modal similarity relationships as soft supervision to boost more intricate alignment.  Extensive experiments validate the effectiveness of our approach, outperforming previous methods by significant margins of at least 3.9\% (T2A) / 6.9\% (A2T) R@1 on the AudioCaps dataset and 2.9\% (T2A) / 5.4\% (A2T) R@1 on the Clotho dataset.
\end{abstract}
\begin{keywords}
audio-text retrieval, multiscale matching, cross-modal similarity
\end{keywords}

%

\section{Introduction}
Audio-text retrieval (ATR) task comprises two subtasks: text-based audio retrieval (T2A) and audio-based text retrieval (A2T). For the former, the target is to retrieve a corresponding audio clip from a collection of candidates given a textual caption, and the latter is just the opposite. This task carries significant application value in domains like search engines and multimedia databases, which gathered considerable attention from the research community \cite{li2019visual,radford2021learning,tan2019lxmert,yao2021filip,liu2022universal}.

In recent years, contrastive learning methods\cite{mei2022metric,deshmukh2022audio,lou2022audio} based on CLAP\cite{elizalde2023clap} framework have been proposed, which aim to  measure the relevance between the audio clips and text captions based on single-vector global representations. 
Despite simplicity, the fine-grained alignment relationships between modalities and the local details are overlooked. 
In addition, relevant studies \cite{lee2018stacked,banerjee2021weaqa} have shown that caption serves as weak supervision, and words in the sentence correspond to specific but unknown frames in the audio. This insight emphasized the need for deeper frame-word alignment understanding to comprehensively explain audio-text matching.
Besides, our preliminary experiments observed that current ATR methods tend to fail when a case has similar but inconsistent descriptions of details in two modalities, which surpasses the capability of single-vector global matching methods to address. 
All these issues prompt us to extract local features and establish fine-grained alignment between text and audio.
Furthermore, 
the utilization of binary contrastive learning labels continues to impose significant constraints on model training, hindering the effective capture of intricate details, including relationships, attributes, and objects.
When matching a text of ``\emph{Car honks three times}" with an audio containing one car honks, it might cause confusion about the amount of honks since conventional cross-modal matching methods mainly emphasize ``\emph{car}" and ``\emph{honk}." 
But it is worth noting that it is much easier to solve such confusions within a single modality of text or audio. 
This means that intra-modality knowledge can also contribute to learning cross-modal alignment in addition to binary contrastive learning labels.
Certain methods resorted to additional training for above information using supplementary data \cite{wang2023one, huang2022masked}. In our case, we aim to extract soft supervisory signals from the intrinsic characteristics of the original data without using any additional data.

To comprehensively address the issues discussed above, this paper proposes a method of multiscale matching driven by cross-modal similarity consistency for ATR. Specifically,  we introduce a multiscale matching architecture, which unfolds in three stages: from local-local to local-global and ultimately to global-global.  Unlike the aggregation methods employed in other approaches using mean or max strategies\cite{zhao2023multi,xin2023improving}, our method allocates different attention to various components at multiple interaction scales.
This approach progressively integrates local information into global matching, encompassing both global and local matching across various scales, enhancing the comprehensive understanding of correlations between modalities.
Furthermore, a novel loss function based on cross-modal similarity consistency (CMSC) is designed for model training. The principle of CMSC can be expressed as follows: {\textit{``If the audio and text descriptions of instances are mapped to the same semantic space, similarity between the representations of two instances should remain consistent, regardless of the modality (i.e. text or audio) used to derive the representation for each instance."}}
\begin{figure*}[t]
  \centering
  \setlength{\abovecaptionskip}{1.cm}
  \includegraphics[width=16cm]{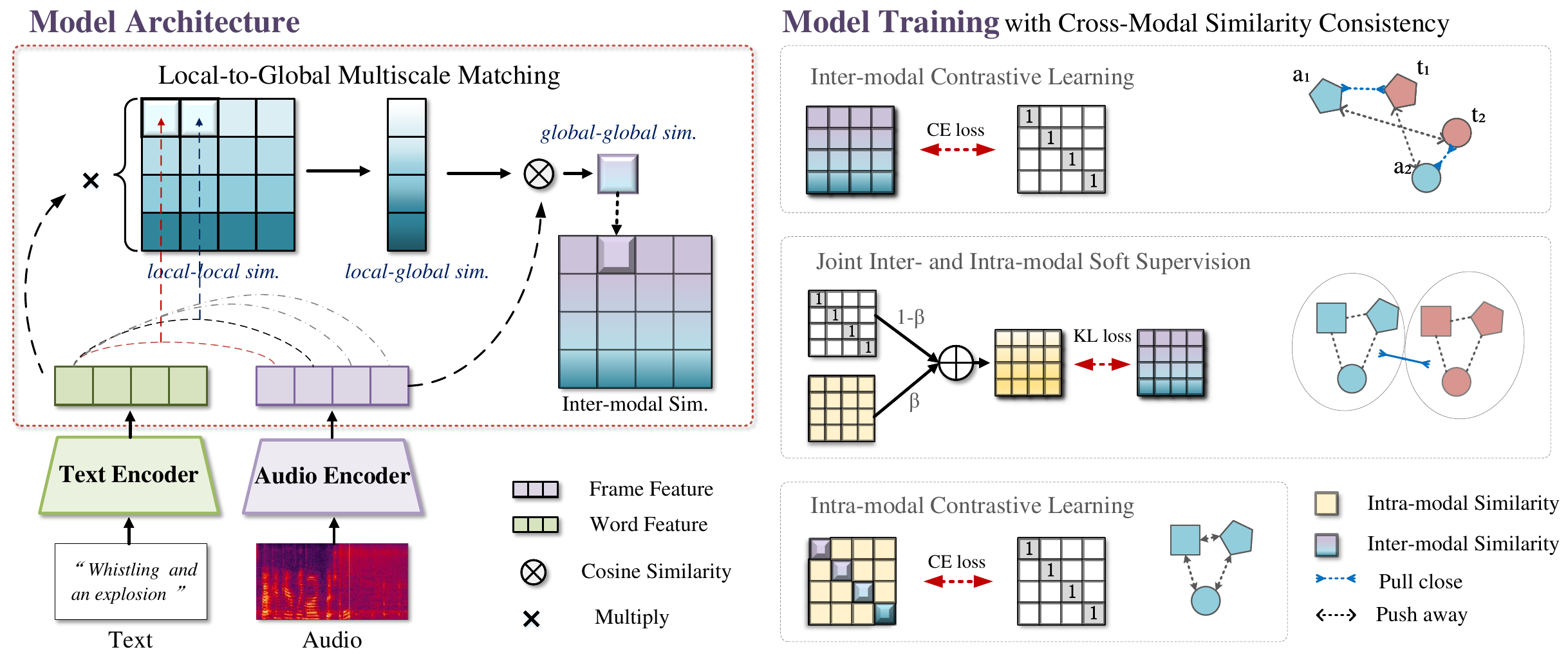}
  \caption{The overall flowchart of our proposed method. The left section represents the model architecture, including the local-to-global multiscale matching (LGMM) module, some of the normalization functions are not displayed in the diagram for simplicity. The right section illustrates the model training, which consists of three loss functions, and their respective impacts are depicted from top to bottom on the right. Color denotes modality (blue for audio and brick red for text), while each shape denotes a data point.}
  \vspace{-3mm}
\end{figure*}
In line with this principle, we introduce a soft supervision loss that measures the consistency between inter-modal and intra-modal similarities. Simultaneously, an intra-modal contrastive learning loss is also utilized to reduce inherent noise (i.e., minor inconsistencies within the same modality). 
To validate the effectiveness of our proposed method, a series of A2T and T2A experiments were carried out on two official benchmarks, AudioCaps\cite{kim2019audiocaps} and Clotho\cite{drossos2020clotho}. Comparative analyses were performed against strong baselines \cite{wu2023large,koepke2022audio,xin2023improving,mei2022metric}, revealing that our approach led to a boost in R@1 performance, achieving an improvement of at least 3.9\% (T2A) and 6.9\% (A2T) on AudioCaps, as 2.9\% (T2A) and 5.4\% (A2T) on Clotho.
\vspace{-2mm}

\section{PROPOSED METHODS}
The overall flowchart of our proposed methodology is depicted in Figure 1.
The left segment illustrates the model architecture, which includes a dual encoder for encoding audio and text, and a subsequent late interaction facilitated by local-to-global multiscale matching. The right segment showcases the model training aspect, the middle of which is driven by intra-modality information supervision that adheres to cross-modal similarity consistency, thereby enabling inter-modality feature training.

Following the conventional setting \cite{wu2023large},  HT-SAT\cite{chen2022hts} is adopted as the audio encoder based on transformer\cite{vaswani2017attention}. For audio clips longer than 10 seconds, we resize the log Mel-spectrograms to a fixed scale. The encoder has an embedding size of 768 and extracts fine-grained representations for ${8\times}$ downsampled frames. BERT \cite{kenton2019bert} is employed to construct the text encoder. Captions are preprocessed to a maximum length of 30 tokens, then audio and text inputs are mapped to a shared semantic feature space of 512 dimensions using two ReLU \cite{glorot2011deep} activated linear layers.

\subsection{Local-to-Global Multiscale Matching}
\label{lgmm}
As depicted in Figure 1, drawing inspiration from the SCAN model \cite{lee2018stacked} and the late interaction concept 
 introduced by COLBERT \cite{khattab2020colbert}, the computation of cross-modal similarity adopts a multiscale framework, progressing from local to global levels.
Taking A2T retrieval as an example, the computation of overall similarity undergoes a three-stage process from \textbf{local-local}, \textbf{local-global} to \textbf{global-global}. Here, we utilize an audio clip as the query and employ text as the context for instance.

Local features ${\bm{F}_a = [\bm{f}_{a,1},\bm{f}_{a,2}... \bm{f}_{a,l_a}]\in\mathbb{R}^{l_a\times dim}}$ from a clip of  audio and ${\bm{F}_t = [\bm{f}_{t,1},\bm{f}_{t,2}... \bm{f}_{t,l_t}]\in\mathbb{R}^{l_t\times dim}}$ from  a sentence of text caption are extracted firstly, where ${l_a}$ and ${l_t}$ respectively represent the number of downsampled frames in the audio and the number of words in the text,  ${dim}$ denotes dimension of the semantic feature space and ${\bm{f}_{a,i}}$ is the local feature of the i-th audio frame, similarly ${\bm{f}_{t,j}}$ is the representation of the j-th word. 

\textbf{Local-Local Interaction}  Firstly, as shown in the left part in Figure 1, the similarity matrix for all frame-word pairs is computed below, 
\begin{equation}
s_{i j}=\bm{f}_{a,i}^\top \bm{f}_{t,j},   i \in[1,l_a], j \in[1, l_t].
\end{equation}
Then the similarity is transformed into a probability distribution ${w_{i j}}$ along the text direction by softmax as  follows,
\begin{equation}
w_{i j}=\frac{\exp \left(\bar{s}_{i j}/\tau_w\right)}{\sum_{j=1}^{l_t} \exp \left(\bar{s}_{i j}/\tau_w \right)},
\end{equation}
where $\bar{s}_{i j}=s_{i j}/ \sqrt{\sum_{i=1}^{l_a}s_{ij}^2}$  and ${\tau_w}$ is a temperature coefficient used to regulate the diversity of the output probability distribution.

\textbf{Local-Global Interaction}  To construct a text-aware audio frame vector ${\bm{v_{a,i}^t}}$, all weighted words are summed
 up with probability distribution attendance as follows,
\begin{equation}
\bm{v}_{a,i}^t=\sum_{j=1}^{l_t} w_{i j} \bm{f}_{t,j} .
\end{equation}
For i-th frame, the new text-aware vector ${\bm{v}_{a,i}^t}$ is compared with the original representation ${\bm{f}_{a,i}}$ using cosine similarity calculation, which contributes to frame-sentence interaction, i.e.,
\begin{equation}
S\left(\bm{f}_{a,i}, \bm{v}_{a,i}^t\right)=\frac{\bm{f}_{a,i}^\top \bm{v}_{a,i}^t}{\left\|\bm{f}_{a,i}\right\|\left\|\bm{v}_{a,i}^t\right\|}.
\end{equation}

\textbf{Global-Global Interaction} In this approach, we calculate the overall (global) similarity score between audio clip A and caption T using LogSumExp pooling (LSE) technique, i.e.,
\begin{equation}
S_{g}(A,T)=\log \left(\sum_{i=1}^{l_a} \exp \left(\lambda S\left(\bm{f}_{a,i}, \bm{v}_{a,i}^t\right)\right)\right)^{\left(1 / \lambda\right)},
\end{equation}
where ${\lambda}$ is a parameter that controls the degree to which the importance of the most relevant features pairs (text-aware vector ${\bm{v}_{a,i}^t}$ and the original audio vector ${\bm{f}_{a,i}}$) should be amplified.

\subsection{Model Training with Cross-Modal Similarity Consistency}
\label{cmsc}
The overall training loss consists of three components: inter-modal contrastive learning, joint inter- and intra-modal soft supervision, and intra-modal contrastive learning. The first component is traditionally used, while the latter two follow the CMSC principle, leveraging intra-modality signals to assist in cross-modal alignment.
Firstly, a batch containing B audio-text pairs $\left\{A_m, T_m\right\}_{m=1}^{B}$ is sampled, and our objective functions are constructed based on ${S_g(A_m,T_n)}$ computed in Section 2.1.

\textbf{Inter-modal Contrastive Learning} To begin with, we employ the binary labels to construct the contrastive learning loss between modalities. This process narrows the gap between positive sample pairs and pushes negative pairs apart, as shown in the right part of Figure1.  Our matching paradigm is trained through optimizing the bidirectional contrastive learning loss, utilizing the normalized temperature-scaled cross-entropy (NT-Xent) loss \cite{chen2020simple} as follows, 
\begin{equation}
\begin{aligned}
\mathcal{L}_{InterC}=-\frac{1}{B} & \left(\sum_{m=1}^B \log \frac{\exp \left(S_g\left(A_m, T_m\right) / \tau\right)}{\sum_{n=1}^B \exp \left(S_g\left(A_m, T_n\right) / \tau\right)}+\right. \\
& \left.\sum_{n=1}^B \log \frac{\exp \left(S_g\left(A_n, T_n\right) / \tau\right)}{\sum_{m=1}^B \exp \left(S_g\left(A_m, T_n\right) / \tau\right)}\right),
\end{aligned}
\label{eq:atc}
\end{equation}
where ${\tau}$ denotes the temperature hyper-parameter.


\textbf{Joint Inter- and Intra-modal Soft Supervision} 
To bring cross-modal similarities closer to the distribution within modality, intra-modal similarities are served as soft labels.  Text-to-text (T2T) and audio-to-audio (A2A) similarities are calculated with the matching method in Section 2.1, denoted as ${S_g(A_m,A_n)}$ and ${S_g(T_m,T_n)}$, respectively.
We perform a weighted combination of binary labels and intra-modal similarities to construct soft labels ${\widehat{S}_g(A_m, A_n)}$ and ${\widehat{S}_g(T_m, T_n)}$ as shown below,
\begin{equation}
\operatorname{\widehat{S}_g}(A_m, A_n)=\beta\operatorname{S_g}(A_m, A_n)+(1-\beta) Y_{m,n} ,
\end{equation}
\begin{equation}
\operatorname {\widehat{S}_g}(T_m, T_n)=\beta \operatorname{S_g}(T_m, T_n)+(1-\beta) Y_{m,n} .
\end{equation}
Here, ${Y_{m,n}}$ represents binary labels indicating whether cross-modal samples match, and ${\beta}$ controls the balance between weak and binary labels. Then, we establish a KL divergence loss to align the cross-modal similarities closer to the designated soft labels i.e.,
\begin{equation}
\begin{gathered}
\mathcal{L}_{Jnt}=\mathcal{D}_{K L}(\operatorname{\widehat{S}_g}(A, A) \mid \operatorname{S_g}(A, T)) / 2 \\
+\mathcal{D}_{K L}(\operatorname{\widehat{S}_g}(T, T) \mid \operatorname{S_g}(T, A)) / 2 ,
\end{gathered}
\label{eq:soft}
\end{equation}
where ${\widehat{S}_g(A,A)}$ denotes the distribution of $\left\{{\widehat{S}_g}(A_m, A_n)\right\}_{m,n}^{B}$, analogous distributions are similarly represented.

\textbf{Intra-modal Contrastive Learning} 
Considering that ${\mathcal{L}_{Jnt}}$ utilizes  intra-modal relationships to guide cross-modal alignment, it is necessary to constrain intra-modal similarity to reduce inherent noise within modalities. An intra-modal contrastive learning loss is established, emphasizing the divergence of same-modality features for negative sample pairs. The formula is as follows,
\begin{equation}
\begin{aligned}
\mathcal{L}_{IntraC}= & -\frac{1}{B} \left(\sum_{m=1}^B \log \frac{\exp \left(S_g\left(A_m, T_m\right) / \tau\right)}{\sum_{n\neq{m}}^B \exp \left(S_g\left(A_m, A_n\right) / \tau\right)} \right. \\
& + \left.\sum_{n=1}^B \log \frac{\exp \left(S_g\left(A_n, T_n\right) / \tau\right)}{\sum_{m\neq{n}}^B \exp \left(S_g\left(T_m, T_n\right) / \tau\right)}\right).
\end{aligned}
\label{eq:NIC}
\end{equation}
\vspace{2mm}
Summarizing the three objective functions mentioned above, we arrive at the overall training loss, which is defined as follows,
\begin{equation}
\mathcal{L}=\mathcal{L}_{InterC}+\mathcal{L}_{Jnt}+\mathcal{L}_{IntraC}.
\label{eq:All}
\end{equation}

\section{Experiments}
\subsection{Datasets}
We conducted text-based audio retrieval and audio-based text retrieval experiments on public datasets: AudioCaps\cite{kim2019audiocaps} and Clotho \cite{drossos2020clotho}.
AudioCaps\cite{kim2019audiocaps} contains 50K short clips extracted from AudioSet\cite{gemmeke2017audio}. The training set consists of 49274 audio clips, each with a sentence of caption. The test set and valid set respectively contain 957 and 494 audio clips, each with five captions.
 Clotho\cite{drossos2020clotho} comprises 5929 audio clips with durations approximately ranging from 15 to 30 seconds. The training, validation, and test sets contain 3839, 1045, and 1045 audio clips, respectively, with each audio clip having five corresponding textual descriptions.

\begin{table*}[]
\begin{center}
\caption{Comparison of Audio-Text Retrieval Performance on Test Sets of AudioCaps and Clotho Datasets.}
\setlength{\tabcolsep}{1.5mm}{
\begin{tabular}{l|llllll|llllll}
\hline
\multicolumn{1}{c|}{\multirow{3}{*}{Model}} & \multicolumn{6}{c|}{AudioCaps}                                                 & \multicolumn{6}{c}{Clotho}                                                     \\ \cline{2-13} 
\multicolumn{1}{c|}{}                       & \multicolumn{3}{c|}{Text-to-Audio}         & \multicolumn{3}{c|}{Audio-to-Text} & \multicolumn{3}{c|}{Text-to-Audio}         & \multicolumn{3}{c}{Audio-to-Text} \\ \cline{2-13} 
\multicolumn{1}{c|}{}                       & R@1   & R@5   & \multicolumn{1}{l|}{R@10}  & R@1        & R@5       & R@10      & R@1   & R@5   & \multicolumn{1}{l|}{R@10}  & R@1       & R@5       & R@10      \\ \hline
MMT \cite{koepke2022audio}                                      & 36.1  & 72.0  & \multicolumn{1}{l|}{84.5}  & 39.6       & 76.8      & 86.7      & 6.5   & 21.6  & \multicolumn{1}{l|}{32.8}  & 6.3       & 22.8      & 33.3      \\
OML \cite{mei2022metric}                                       & 33.9  & 69.7  & \multicolumn{1}{l|}{82.6}  & 39.4       & 72.0      & 83.9      & 14.4  & 36.6  & \multicolumn{1}{l|}{49.9}  & 16.2      & 37.5      & 50.2      \\
TAP\cite{xin2023improving}                                         & 36.1  & 72.0  & \multicolumn{1}{l|}{85.2}  & 41.3       & 75.5      & 86.1      & 16.2  & 39.2  & \multicolumn{1}{l|}{50.8}  & 17.6      & 39.6      & 51.4      \\
HTSAT-CLAP\cite{wu2023large}                                & 36.7  & 70.9  & \multicolumn{1}{l|}{83.2}  & 45.3       & 78.0      & 87.7      & 12.0  & 31.6  & \multicolumn{1}{l|}{43.9}  & 15.7      & 36.9      & 51.3      \\ \hline
Ours                              & \textbf{40.6} & \textbf{74.5} & \multicolumn{1}{l|}{\textbf{86.0}} & \textbf{52.2}      & \textbf{82.5}     & \textbf{91.3}     & \textbf{19.1} & \textbf{44.2} & \multicolumn{1}{l|}{\textbf{56.3}} & \textbf{23.0}     & \textbf{48.5}     & \textbf{61.2}     \\ \hline
\end{tabular}}
\vspace{-4mm}
\end{center}
\end{table*}

\subsection{Training and Metrics}
Our models underwent training for 40 epochs, employing a batch size of 128 and a learning rate of ${5\times{10^{-5}}}$ with optimization through the Adam optimizer\cite{kingma2014adam}.
Regarding hyperparameters, the values of the temperature of inter-and intra-modal contrastive learning loss ${\tau}$ were set to 0.07. For our multiscale matching framework, the values of the temperature of softmax on the similarity matrix ${\tau_w}$ and the inversed temperature ${\lambda}$ were set to 0.25 and 10, respectively. In terms of the proposed joint inter- and intra-modal soft supervision, we designated ${\beta}$ as 0.3.

Following the setting of the benchmarks, the assessment of audio-text retrieval performance across models was established upon the metrics R@1, R@5, and R@10.

\subsection{Evaluation Results}
This section presents a comprehensive series of experiments conducted on two datasets to validate the effectiveness of our model.

\textbf{Comparison with Other Works }
We compared our work with other approaches derived from the CLAP\cite{elizalde2023clap} architecture( i.e. OML\cite{mei2022metric}, TAP\cite{xin2023improving}, HTSAT-CLAP \cite{wu2023large}), as well as fine-grained framework MMT.
The performance comparison presented in Table1 demonstrates that our model outperforms those coarse-grained matching works with a single vector (Line 2-4), showcasing an impressive improvement of 3.9\% and 2.9\% on R@1 in terms of text-to-audio, 6.9\% and 5.4\% in terms of audio-to-text on AudioCaps and Clotho, respectively. In comparison to the fine-grained matching achieved by the full cross-modal interaction in MMT\cite{koepke2022audio}, our model particularly demonstrates a lead of 12.6\% and 16.7\% in terms of R@1 for T2A and A2T on the Clotho dataset.
\begin{table}[]
\caption{Performances on Different Fine-grained Matching Approaches.}
\setlength{\tabcolsep}{1.5mm}
\begin{tabular}{l|llllll}
\hline
\multicolumn{1}{c|}{\multirow{2}{*}{Model}} & \multicolumn{6}{c}{AudioCaps}                                                                                     \\ \cline{2-7} 
\multicolumn{1}{c|}{}                       & \multicolumn{3}{c|}{Text-to-Audio}                                 & \multicolumn{3}{c}{Audio-to-Text}             \\ \hline
\multicolumn{1}{c|}{}                       & R@1           & R@5           & \multicolumn{1}{l|}{R@10}          & R@1           & R@5           & R@10          \\ \cline{2-7} 
Max-Mean                                & 38.9          & 73.7          & \multicolumn{1}{l|}{\textbf{85.1}} & 47.8          & 80.2          & 90.1          \\
Max-Max                                & 37.3          & 72.6         & \multicolumn{1}{l|}{84.5} & 47.8          & 79.2          & 89.7          \\
Mean-Mean                                & 38.6         & 74.0          & \multicolumn{1}{l|}{85.0} & 48.8          & 80.7          & 89.8          \\
Mean-Max                                & 37.6         & 73.2          & \multicolumn{1}{l|}{84.9} & 49.6          & 79.2          & 89.6          \\
Ours (LGMM)                                    & \textbf{40.0} & \textbf{74.1} & \multicolumn{1}{l|}{84.9}          & \textbf{51.4} & \textbf{81.2} & \textbf{90.9} \\ \hline
\end{tabular}
\vspace{-3mm}
\end{table}

\textbf{Comparison with Other Fine-grained Matching Modes}
To validate our local-to-global multiscale matching (LGMM) module, we compared it to traditional fine-grained matching mode Max-Mean and variations (Max-Max, Mean-Mean, and Mean-Max). 
At this point, the LGMM module is trained using only the fundamental loss ${\mathcal{L}_{InterC}}$ as described in Eq.\ref{eq:atc}.
Table2 shows that Max-Max performed the worst due to its sole focus on the most prominent features, neglecting the influence of other elements. Similarly, methods using Mean mechanisms at one stage lacked substantial improvements because they treated all components equally. In contrast, our approach assigns varying weights to components in multiscale modules during gradual aggregation, achieving significant performance improvement. 

\textbf{Ablation Studies on Loss Functions}
We evaluated T2A and A2T retrieval using the objective functions outlined in Section 2.2. Analyzing Table3, removing only ${\mathcal{L}_{Jnt}}$ results in a significant drop in all metrics, confirming enhancement gained from the weak supervision provided by intra-modal relationships. Similarly, the removal of ${\mathcal{L}_{IntraC}}$ leads to a more drastic overall decline in metrics, indicating it enhances performance by pushing apart negative sample features within the same modality. This reduces intra-modal noise and indirectly influences cross-modal alignment. However, removing ${\mathcal{L}_{IntraC}}$ actually slightly improves R@1 for A2T on AudioCaps, since  cross-modal relationships tend to unconditionally be closer to intra-modal relationships in the absence of ${\mathcal{L}_{IntraC}}$. In this case, the inherent similarity relationships within A2T closely resemble those within A2A, resulting in improved R@1 performance. Significantly,  ${\mathcal{L}-\mathcal{L}_{IntraC}-\mathcal{L}_{Jnt}}$ performs the worst across all benchmarks, demonstrating the effectiveness of the CMSC method in ATR task.

\vspace{-2mm}
\subsection{Case Study}
To further verify the effectiveness of our fine-grained alignment method, the local-global (i.e.,word-to-clip) similarities on three audio clips towards one caption are visualized in the Figure2. ${A_0}$ represents the positive audio sample corresponding to the query caption ``\emph{A car engine revs producing a room and a whine}", while ${A_1}$ and ${A_2}$ serve as hard negatives, with corresponding textual descriptions being ``\emph{A powerful engine revs as it idles}" and ``\emph{Low humming of an idling and accelerating engine}", respectively. All three segments describe car engine sounds, but ``\emph{a whine}"  (a sharp sound) only appeared in ${A_0}$.
Figure2(a) reveals that our approach distinctly shows lower similarity scores for ${A_1}$/${A_2}$ with the phrase ``\emph{a whine}", suggesting that the sharp sound is nearly absent. In contrast, ${A_0}$ has a notably higher probability of containing this, aligning with the retrieval fact.
Figure 2(b) depicts the Max-Mean method, which selects the most salient frame and averages similarity with all words, exhibiting minimal variation in word-to-clip similarity. Consequently, its ability to discern the phrase ``\emph{a whine}" is weak. The similarity ranking with the entire sentence using this method is ${A_2>A_1>A_0}$, failing to retrieve the ground truth ${A_0}$.


\begin{table}[]
\caption{Experimental Results from Ablation Studies on Loss Functions.}
\setlength{\tabcolsep}{1.5mm}
\begin{tabular}{l|llllll}
\hline
\multicolumn{1}{c|}{\multirow{2}{*}{model}} & \multicolumn{3}{c|}{Text-to-Audio}      & \multicolumn{3}{c}{Audio-to-Text} \\ \cline{2-7} 
\multicolumn{1}{c|}{}                       & R@1  & R@5  & \multicolumn{1}{l|}{R@10} & R@1       & R@5       & R@10      \\ \hline
\multicolumn{7}{c}{AudioCaps}                                                                                            \\ \hline
\multicolumn{1}{l|}{${\bm{\mathcal{L}}}$}      & ${\bm{40.6}}$ & ${\bm{74.5}}$ & \multicolumn{1}{l|}{${\bm{86.0}}$} & 52.2      & ${\bm{82.5}}$      & 91.3      \\ 
\multicolumn{1}{l|}{${-\mathcal{L}_{Jnt}}$}                & 40.2 & 74.3 & \multicolumn{1}{l|}{85.8} & 51.9      & 82.3      & ${\bm{91.5}}$      \\ 
\multicolumn{1}{l|}{${-\mathcal{L}_{IntraC}}$}                & 39.7 & 74.3 & \multicolumn{1}{l|}{85.5} & ${\bm{52.5}}$      & 78.4      & 89.9      \\
\multicolumn{1}{l|}{${-\mathcal{L}_{IntraC}-\mathcal{L}_{Jnt}}$}                & 40.0 & 74.1 & \multicolumn{1}{l|}{84.9} & 51.4      & 81.2      & 90.9      \\  \hline
\multicolumn{7}{c}{Clotho} \\ \hline
\multicolumn{1}{l|}{${\bm{\mathcal{L}}}$}                & ${\bm{19.1}}$ & ${\bm{44.2}}$ & \multicolumn{1}{l|}{56.3} & ${\bm{23.0}}$      & 48.5      & ${\bm{61.2}}$    \\
\multicolumn{1}{l|}{${-\mathcal{L}_{Jnt}}$}                & 18.9 & 43.9 & \multicolumn{1}{l|}{56.5} & 22.8      & ${\bm{48.8}}$      & 61.0     \\
\multicolumn{1}{l|}{${-\mathcal{L}_{IntraC}}$}                & 17.4 & 42.1 & \multicolumn{1}{l|}{${\bm{56.8}}$} & 21.1      & 46.3      & 60.8      \\
\multicolumn{1}{l|}{${-\mathcal{L}_{IntraC}-\mathcal{L}_{Jnt}}$}                & 17.2 & 43.5 & \multicolumn{1}{l|}{56.5} & 20.1      & 45.8      & 58.9     \\
\hline
\end{tabular}
\end{table}
\vspace{-2mm}

\begin{figure}[t]
  \centering
  \setlength{\abovecaptionskip}{1.cm}
  \includegraphics[width=8cm]{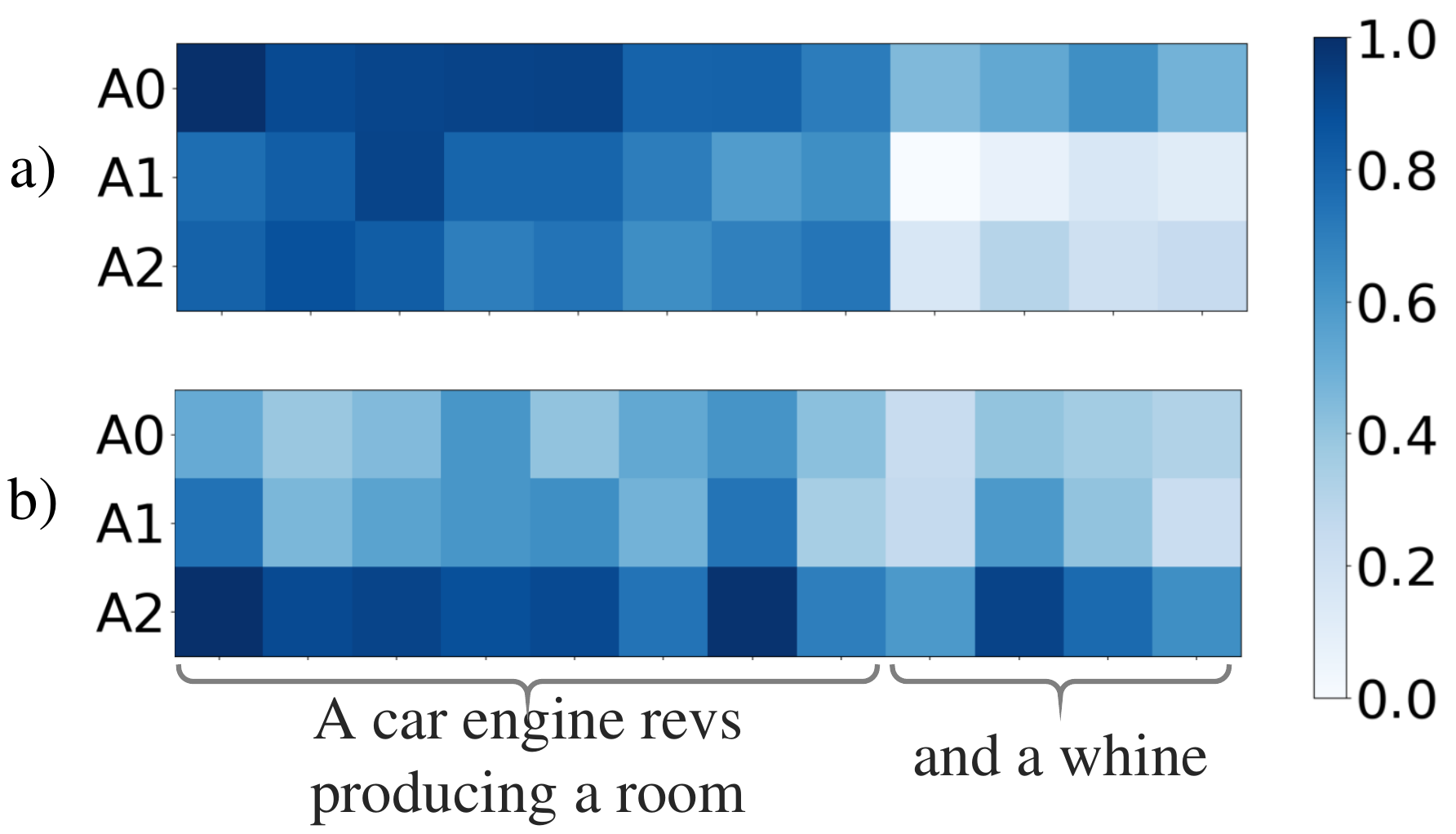}
  \caption{Visualization of word-to-clip similarities given by the (a) Ours (LGMM) model and the (b) Max-Mean model in Table2.}
\end{figure}

\section{Conclusion}
This paper presents a novel local-to-global multiscale matching approach designed to address the limitations of coarse-grained retrieval methods that overlook detailed alignment in audio-text retrieval tasks. Experimental results demonstrate that our framework establishes finer and more intricate alignment relationships, achieving significant improvements in performance without using additional data compared with previous methods. To tackle the challenge posed by binary labels that only indicate whether cross-modal pairs match, we leverage the inherent relative correlations within each modality. This strategic utilization of intra-modal relationships aids in constructing inter-modal relationships, contributing to further performance enhancements. In the future, the utilization of fine-grained features represented by the audio event vectors sourced from sound event detection will be explored as part of our research.


\vfill\pagebreak
\bibliographystyle{IEEEbib}
\bibliography{strings,refs}

\end{document}